\documentclass{ws-procs9x6}

\begin{document}

\title{Gravitational Waves from Compact Sources\
\footnote{\uppercase{P}roceedings of the 5th
\uppercase{I}nternational \uppercase{W}orkshop ``\uppercase{N}ew
\uppercase{W}orlds in \uppercase{A}stroparticle
\uppercase{P}hysics'' }}

\author{K.~D. KOKKOTAS and N. STERGIOULAS\
 \footnote{\uppercase{T}his work has been supported by
the \uppercase{EU} program \uppercase{ILIAS(ENTA}p\uppercase{P}) and
the \uppercase{GSRT} program \uppercase{H}eracleitus.}}

\address{Department of Physics, \\
Aristotle University of Thessaloniki, \\
54124 Thessaloniki, Greece.\\
E-mail: \ kokkotas@auth.gr,  \  niksterg@auth.gr}

\maketitle

\abstracts{We review sources of high-frequency
gravitational waves, summarizing our current understanding of
emission mechanisms, expected amplitudes and event
rates. The most promising sources are gravitational collapse
(formation of black holes or neutron stars) and subsequent
ringing of the compact star, secular or dynamical rotational
instabilities and high-mass compact objects formed through
the merger of binary neutron stars. Significant and unique
information for the various stages of the collapse, the structure
of protoneutron stars and the high density equation of state of
compact objects can be drawn from careful study of gravitational
wave signals. }




\section{Introduction}

The new generation of gravitational wave (GW) detectors is already
collecting data, improving the achieved sensitivity by at least one
order of magnitude compared to operating resonant bar detectors.
Broadband GW detectors are sensitive to frequencies between 50 and a
few hundred Hz. In their advanced stage, the current GW detectors
will have a broader bandwidth but will still not be sensitive
enough to frequencies above 500 to 600Hz. Nevertheless, improved
sensitivity can be achieved at high frequencies through
narrow-band operation \cite{ThorneCutler,GEO600}. In addition, there
are proposals for constructing wide band resonant detectors in the kHz
band \cite{Cerdonio}.

In this short review we will discuss some of the sources that are in
the high frequency band ($\gtrsim 500-600$Hz), where the
currently operating interferometers are sensitive enough only if they
are narrowbanded. Since there exist a variety of GW sources with very
interesting physics to be explored, this high-frequency window
deserves special attention. If either resonant or narrow-band
interferometers achieve the required sensitivity, a plethora
of unique information can be deduced from detected signals.

\section{Gravitational collapse}

{\it Core collapse.} One of the most spectacular astrophysical events
is the core collapse of massive stars, leading to the
formation of a neutron star (NS) or a black hole (BH).
The outcome of core collapse depends sensitively on several
factors: mass, angular momentum and metallicity of progenitor,
existence of a binary companion, high-density equation of
state, neutrino emission, magnetic fields, etc. Partial
understanding of each of the above factors is
emerging, but a complete and consistent theory for core
collapse is still years away.

Roughly speaking, isolated stars more massive than $\sim
8-10M_\odot$ end in core collapse and  $\sim 90\%$ of them are
stars with masses $\sim
8-20M_\odot$. After core bounce, most of the material is ejected and
if the progenitor star has a mass $M\lesssim 20M_\odot$ a
neutron star is left behind. On the other hand, if $M\gtrsim
20M_\odot$ fall-back accretion increases the mass of the
formed proton-neutron star (PNS), pushing it above the maximum
mass limit, which results in the formation of a black hole.
Furthermore, if the progenitor star has a mass of roughly $M\gtrsim
45M_\odot$, no supernova explosion is launched and the star collapses
directly to a BH\cite{Fryer99}.

The above picture is, of course, greatly simplified. In reality,
the metallicity of the progenitor, the angular momentum of the
pre-collapse core and the presence of a binary companion will
decisively influence the outcome of core collapse\cite{Fryer2003}.
Very massive stars lose mass through strong
stellar winds. The mass loss rate is sensitive to the metallicity
of the star and can be very high, allowing a 60 $M_\odot$ star to
leave behind a typical 1.4 $M_\odot$ neutron star, instead of a much
more massive black hole. Since mass-loss is a complex phenomenon,
current observations are still insufficient to constrain the
different
possible outcomes of core collapse for stars with $M\gtrsim
45M_\odot$. Through mass transfer or common envelope
episodes, a binary companion can cause a massive star to enter the
rapid mass-losing (Wolf-Rayet) phase earlier, making wind mass loss
more effective. Roughly half of all stars are in suffiently close
binaries that binary interactions must be taken into account when
trying to predict to outcome of core collapse.
Rotation influences the collapse by changing
dramatically the properties of the convective region above the
proto-neutron star core. Centrifugal forces slow down infalling
material in the equatorial region compared to materiall falling in
along the polar axis, yielding a weaker bounce. This
asymmetry between equator and poles also strongly influences the
neutrino emmission and the revival of the stalled shock by neutrinos
\cite{Fryer2004,Burrows2005}.

The supernova event rate is 1-2 per century per galaxy
\cite{Cappellaro} and about 5-40\% of them produce BHs in
delayed collapse (through fall-back accretion), or direct
collapse\cite{FryerKalogera}.

\vskip 0.4cm

\noindent {\it Initial rotation rates.} Of considerable importance
is the {\em initial rotation rate} of proto-neutron stars, since (as
will be detailed in the next sections) most mechanisms for
emission of detectable gravitational waves from compact
objects require very rapid rotation at birth (rotational
periods of the order of a few milliseconds or less).
Since most massive stars have non-negligible rotation rates
(some even rotate near their break-up limit), simple conservation
of angular momentum would suggest a proto-neutron star
to be strongly differentially rotating with very high rotation
rates and this picture is supported by numerical simulations
of rotating core collapse\cite{FryerHeger,DFM2002b}.

On the other hand, observationally, we know of
pulsars (that have been associated with a supernova remnant)
rotating only as fast as with a 16ms period, which suggest
a period of at least several milliseconds at birth\cite{Marschall98}.
A possible explanation for this discrepancy is that the
pre-supernova core has been slowed down by e.g. magnetic
torques\cite{Spruit2002}. The most recent evolutionary models of
rotating core-collapse
progenitors (including the effect of magnetic torques) suggest
that progenitors of neutron stars with typical masses of
$\sim 1.4 M_\odot$ are indeed slowly rotating, producing
remnants with periods at birth of the order of 10 ms. Nevertheless,
the same study finds that very massive progenitors evolve so
rapidly that the angular momentum transfer out of the core, by
magnetic torques, is diminished, yielding heavy proto-neutron
stars with rotational periods of the order of a few milliseconds
(provided mass-loss takes place to prevent black hole
formation)\cite{Heger05}. Binary interactions can also accelerate some
phases of progenitor
evolution, allowing for fast initial spins\cite{Pods04,Petrovic05}.
If the magnetic torques operate efficiently, it is clear that rapid
rotation at birth will have an event rate much smaller than the usual
galactic supernova rate. Still, strong emission mechanisms (e.g.
bar-mode instabilities) could yield detectable signals at acceptable
event rates.

Other ways to form a rapidly rotating proto-neutron star would
be through {\it fall-back accretion}\cite{Watts02},
through the {\it accretion-induced collapse of a white
dwarf}\cite{Hillebrandt1984,Liu2001,Fryer2002,Yoon05} or
through the merger of binary white dwarfs in globular
clusters\cite{Middleditch04}. It is also relevant to take into account
current gamma-ray-burst models. The {\it collapsar}\cite{Woosley93}
model requires high rotation rates of a proto-black
hole\cite{Petrovic05}. In addition, a possible formation scenario for
magnetars involves a rapidly rotating protoneutron star formed through
the collape of a very massive progenitor and some observational
evidence is already emerging\cite{Gaensler05}.

\vskip 0.4cm

\noindent{\it Gravitational wave emission.} Gravitational waves
from core collapse have a rich spectrum, reflecting the
various stages of this event. The initial signal is emitted due
to the {\it changing axisymmetric quadrupole moment} during collapse.
In the case of neutron star formation, the quadrupole moment
typically becomes larger, as the core spins up during contraction.
In contrast, when a rapidly rotating neutron star collapses to
form a Kerr black hole, the axisymmetric quadrupole moment
first increases but is finally reduced by a large factor when the
black hole is formed.

A second part of the gravitational wave signal is produced
when gravitational collapse is halted by the stiffening
of the equation of state above nuclear densities and the core bounces,
driving an outwards moving shock. The dense fluid undergoes motions
with relativistic speeds ($v/c\sim 0.2-0.4$) and a rapidly rotating
proto-neutron star thus oscillates in several of its
axisymmetric {\it normal modes of oscillation}.
This quasi-periodic part of the signal could last for hundreds of
oscillation periods, before being effectively damped.
If, instead, a black hole is directly formed, then black hole
quasinormal modes are excited, lasting for only a few
oscillation periods. A combination of neutron star and black
hole oscillations will appear if the proton-neutron star is not stable
but collapses to a black hole.

In a rotating proto-neutron star, nonaxisymmetric processes can
yield additional types of gravitational wave signals. Such processes
are {\it dynamical instabilities}, {\it secular
gravitational-wave driven instabilities} or {\it convection} inside
the proto-neutron star and in its surrounding hot envelope.
{\it Anisotropic neutrino emission}  is accompanied by a gravitational
wave signal. {\it Nonaxisymmetries} could
already be present in the pre-collapse core and become amplified
during collapse\cite{Fryer2004b}. Furthermore, if there
is persistent fall-back accretion onto a proto-neutron
star or black hole, these can be brought into {\it ringing}.

Below, we discuss in more detail those processes which result in
high frequency gravitational radiation.

\subsection{Neutron star formation}
Core collapse as a potential source of GWs has been studied for more
than three decades (some of the most recent simulations can be
found in
\cite{Zwerger1997,Rampp1998,Fryer2002,DFM2002b,Ott2003,Kotake2003,Mueller2004,Shibata04,Shibata05,Sekiguchi05}). The main differences between the various
studies are the
progenitor models (slowly or rapidly rotating), equation of
state (polytropic or realistic), gravity (Newtonian
or relativistic) and neutrino emission (simple, sophisticated or
no treatment). In general, the gravitational wave signal from
neutron star formation is divided into a core bounce signal, a
signal due to convective motions and a signal due to anisotropic
neutrino emission.

\vskip 0.4cm

\noindent {\it Core bounce signal.} The core bounce signal is
produced due to rotational flattening and excitation of normal modes
of oscillations, the main contributions coming from the axisymmetric
quadrupole ($l=2$) and quasi-radial ($l=0$) modes (the latter
radiating through its rotationally acquired $l=2$ piece). If
detected, such signals will be a unique probe for the high-density
EOS of neutron stars\cite{AK1996,AK1998}. The strength of this
signal is sensitive to the available angular momentum in the
progenitor core. If the progenitor core is rapidly rotating, then
core bounce signals from Galactic supernovae ($d\sim 10$kpc) are
detectable even with the initial LIGO/Virgo sensitivity at
frequencies $\lesssim$1kHz. In the best-case scenario, advanced LIGO
could detect signals from distances of 1Mpc, but not from the Virgo
cluster ($\sim$15Mpc), where the event rate would be high. The
typical GW amplitude from 2D numerical simulations
\cite{DFM2002b,Ott2003} for an observer located in the equatorial
plane of the source is\cite{Shibata04}
\begin{equation}
h\approx 9 \times 10^{-21}\varepsilon \left(\frac{ 10 {\rm
kpc}}{d}\right)
\end{equation}
where $\varepsilon \sim 1$ is the normalized GW amplitude. For
such rapidly rotating initial models, the total energy radiated
in GWs during the collapse is $\lesssim
10^{-6}-10^{-8} M_\odot c^2$.
If, on the other hand, progenitor cores are slowly rotating (due to
e.g. magnetic torques\cite{Spruit2002}), then the signal strength is
significantly reduced, but, in the best case, is still within reach of
advanced LIGO for galactic sources.

Normal mode oscillations, if excited in an equilibrium star at a
small to moderate amplitude, would last for hundreds to thousands of
oscillation periods, being damped only slowly by gravitational wave
emission or viscosity. However, the proto-neutron star immediately
after core bounce has a very different structure than a cold
equilibrium star. It has a high internal temperature and is
surrounded but an extended, hot envelope. Nonlinear oscillations
excited in the core after bounce can penetrate into the hot
envelope. Through this damping mechanism, the normal mode
oscillations are damped on a much shorter timescale (on the order of
ten oscillation periods), which is typically seen in the core
collapse simulations mentioned above.

\vskip 0.4cm

\noindent {\it Convection signal.}
The post-shock region surrounding a proto-neutron star is
convectively unstable to both low-mode and high-mode
convection. Neutrino emission also drives convection in this
region. The most realistic 2D simulations of core collapse
to date\cite{Mueller2004} have shown that the gravitational wave
signal from convection significantly exceeds the core bounce
signal for slowly rotating progenitors, being detectable with
advanced LIGO for galactic sources, and is detectable even
for nonrotating collapse. For slowly rotating collapse, there
is a detectable part of the signal in the high-frequency
range of 700Hz-1kHz, originating from convective motions that
dominate around 200ms after core bounce. Thus, if both core
a bounce signal and a convection signal would be detected in
the same frequency range, these would be well separated in time.

\vskip 0.4cm

\noindent {\it Neutrino signal.}
In many simulations the gravitational wave signature of
anisotropic neutrino emission has also been
considered\cite{Epstein78,Burrows96,Mueller97}. This type
of signal can be detectable by advanced LIGO
for galactic sources, but the main contribution is at
low frequencies for a slowly rotating progenitor\cite{Mueller2004}.
 For rapidly rotating progenitors, stronger contributions at
high frequencies could be present, but would probably be
burried within the high-frequency convection signal.

\vskip 0.4cm
Numerical simulations of neutron star formation have gone
a long way, but a fully consistent 3D simulation including
relativistic gravity, neutrino emission and magnetic fields
is still missing. The combined treatment of these effects might not
change the above estimations by orders of magnitude but it will
provide more conclusive answers. There
are also issues that need to be understood such as pulsar kicks
(velocities exceeding 1000 km/s) which suggest that in a
fraction of newly-born NSs (and probably BHs) the formation
process may be strongly asymmetric\cite{Hoef02}.
Better treatment of the microphysics and construction of accurate
progenitor models for the angular momentum distributions are needed.
All these issues are under investigation by many groups.

\subsection{Neutron star ringing through fall-back accretion}

A possible mechanism for the excitation of oscillations in a
proto-neutron star after core bounce is the fall-back accretion of
material that has not been expelled by the revived supernova shock.
The isotropy of this material is expected to be broken due to e.g.
rotation or nonaxisymmetric convective motions, thus a large number
of oscillation modes will be excited as this material falls back
onto the neutron star. This process is, of course, complex and the
detectability of gravitational waves from these oscillations will
depend on several factors, such as the fall-back accretion rate, the
degree of asymmetry of the fall-back material the structure of the
proto-neutron star envelope, the presence of magnetic fields etc.

Recently, the ringing of a neutron star through fall-back accretion
has been modeled through relativistic 2D nonlinear hydrodynamical
simulations\cite{Nagar04}. Quadrupolar shells of matter were
accreted on a static neutron star (in the approximation that the
background spacetime remains unchanged). Gravitational waves were
then extracted through the Zerilli-Moncrief formalism. The
gravitational wave signal from such a process comprises a {\it
narrow peak} at the $l=2$ normal mode frequency of the neutron star
and a very {\it broad peak}, featuring {\it interference fringes},
centered at a much higher frequency. Since the frequency of the
broad peak is still too low to be identified with a $w$-mode of the
star, the interpretation for this part of the signal is that it is
related to the motion of the fluid shell and the reflection of the
gravitational-wave pulse from this motion in the external Zerilli
potential, which also creates the interference fringes. The
accretion of a quadrupolar shell containing 1\% of the mass of the
star releases gravitational waves with a total energy similar to the
energy emitted immediately after core bounce. It is thus interesting
to consider this mechanism in more detail, since the excitation of
the normal modes in the neutron star happens when the star has
already cooled somewhat (compared to the proto-neutron star
immediately after core bounce) which simplifies the identification
of observed oscillations with normal modes of cold neutron star
models. The formation of a dense torus as as a result of stellar
gravitational collapse, binary neutron star merger or disruption.
Such a system either becomes unstable to the runaway instability or
exhibit a regular oscillatory behavior, resulting in a
quasi-periodic variation of the accretion rate as well as of the
mass quadrupole suggesting a new sources of potentially detectable
gravitational waves\cite{Zanotti2003}.

\subsection{Black hole formation}

The gravitational-wave emission from the formation of a Kerr BH is
a sum of two signals: the {\it collapse signal} and the {\it BH
ringing}. The collapse signal is produced due to the changing
multipole moments of the spacetime during the transition from a
rotating iron core or proto-neutron star to a Kerr BH. A uniformly
rotating neutron star has an axisymmetric quadrupole moment given
by\cite{Laarakkers99}
\begin{equation}
Q=-a\frac{J^2}{M}
\end{equation}
where $a$ depends on the equation of state and is in the range
of $2-8$ for 1.4$M_\odot$ models. This is several times larger in
magnitude than the corresponding qudrupole moment of
a Kerr black hole ($a=1$). Thus, the {\it reduction} of the
axisymmetric
quadrupole moment is the main source of the collapse signal. Once
the BH is formed, it continues to oscillate in its axisymmetric
$l=2$ QNM, until all oscillation energy is radiated away and the
stationary Kerr limit is approached.

The numerical study of rotating collapse to BHs was pioneered by
Nakamura\cite{Nakamura81}
 but first waveforms and gravitational-wave estimates were obtained
by Stark and Piran\cite{Stark87}
. These simulations we performed in 2D, using approximate initial data
(essentially a spherical star to which angular momentum was
artificially added). A new 3D computation of the gravitational wave
emission from the collapse of unstable uniformly rotating
relativistic polytropes to Kerr BHs\cite{Baiotti05}
finds that the energy emitted is
\begin{equation}
\Delta E \sim 1.5 \times 10^{-6} (M/M_\odot),
\end{equation}
significantly less than the result of Stark and Piran.
Still, the collapse of an unstable 2$M_\odot$ rapidly rotating
neutron star leads to a characteristic gravitational-wave amplitude
$h_c \sim 3\times 10^{-21}$, at a frequency of $\sim 5.5$kHz, for an
event at 10kpc. Emission is mainly through the "+" polarization, with
the "$\times$" polarization being an order of magnitude weaker.

Whether a BH forms promptly after collapse or a delayed collapse
takes place depends sensitively on a number of factors, such as the
progenitor mass and angular momentum and the high-density EOS. The
most detailed investigation of the influence of these factors on
the outcome of collapse has been presented recently
in\cite{Sekiguchi05}, where it was found that shock formation
increases the threshold for
black hole formation by $\sim 20-40\%$, while rotation results in an
increase of at most 25\%.

\subsection{Black hole ringing through fall-back or hyper-accretion}

{\it Single events.}
A black hole can form after core collapse, if fall-back
accretion increases the mass of the proto-neutron star above the
maximum mass allowed by axisymmetric stability. Material falling back
after the black hole is formed excites the black hole quasi-normal
modes of oscillation. If, on the other hand, the black hole is formed
directly through core collapse (without a core bounce taking place)
then most of the material of the progenitor star is accreted at
very high rates ($\sim 1-2M_\odot$/s) into the hole.
In such {\it hyper-accretion} the black hole's quasi-normal modes
(QNM) can be excited for as long as the process lasts and
until the black hole becomes stationary. Typical frequencies of the
emitted GWs are in the range 1-3kHz for $\sim 3-10 M_\odot$ BHs.

The frequency and the damping time of the oscillations for the
$l=m=2$
mode can be estimated via the relations \cite{Echeverria}
\begin{eqnarray}
\sigma&\approx& 3.2 {\rm kHz} \
M_{10}^{-1}\left[1-0.63(1-a/M)^{3/10}\right] \\
 Q&=&\pi \sigma
\tau \approx 2\left(1-a\right)^{-9/20} \label{bhqnm}
\end{eqnarray}
These relations together with similar ones either for the 2nd QNM or
the $l=2$, $m=0$ mode can uniquely determine the mass $M$ and
angular momentum parameter $a$ of the BH if the frequency and the
damping time of the signal have been accurately extracted
\cite{Finn,Nakano2003,Dryer}. The amplitude of the ring-down waves
depends on the BH's initial distortion, i.e. on the nonaxisymmetry
of the blobs or shells of matter falling into the BH. If matter of
mass $\mu$ falls into a BH of mass $M$, then the gravitational wave
energy is roughly
\begin{equation}
\Delta E \gtrsim
\epsilon \mu c^2(\mu/M)
\end{equation}
 where $\epsilon$ is related
to the degree of asymmetry and could be $\epsilon
\gtrsim 0.01$ \cite{Davis1971}. This
leads to an effective GW amplitude
\begin{equation}
h_{\rm eff}\approx 2\times 10^{-21}\left(\frac{\epsilon}{0.01}
\right)\left(\frac{10 {\rm Mpc}}{d}\right)\left( \frac{\mu}{M_\odot}\right)
\end{equation}

\vskip 0.4cm

\noindent {\it Resonant driving.}
If hyper-accretion proceeds through an accretion disk around a rapidly
spinning Kerr BH, then the matter near the marginally bound
orbit radius can become unstable to the magnetorotational (MRI)
instability, leading to the formation of large-scale
asymmetries\cite{Araya04}. Under certain conditions, {\it resonant
driving} of the BH QNMs
could take place. Such a continuous signal could be integrated,
yielding a much larger signal to noise ratio than a single event.
For a 15$M_\odot$ nearly maximal Kerr BH created at 27Mpc the
integrated signal becomes detectable by LIGO II at a frequency
of $\sim 1600$Hz, especially if narrow-banding is used\cite{Araya04}.

\section{Rotational instabilities}

If proto-neutron stars rotate rapidly, nonaxisymmetric {\it dynamical
instabilities} can develop. These arise from non-axisymmetric
perturbations having angular dependence $e^{i
m\phi}$ and are of two different types: the {\it classical bar-mode}
instability and the more recently discovered {\it low-$T/|W|$
bar-mode} and {\it one-armed spiral} instabilities, which appear to be
associated to the presence of
corotation points. Another class of nonaxisymmetric instabilities are
{\it secular instabilities}, driven by dissipative effects, such as
fluid viscosity or gravitational radiation.

\subsection{Dynamical instabilities}

{\it Classical bar-mode instability.}
The classical $m=2$ bar-mode instability is excited in Newtonian
stars when the ratio $\beta=T/|W|$ of the rotational kinetic energy
$T$ to the gravitational binding energy $|W|$ is larger than
$\beta_{\rm dyn}=0.27$.
The instability grows on a dynamical time scale (the time
that a sound wave needs to travel across the star) which is about one
rotational period and may last from 1 to 100 rotations depending on
the degree of differential rotation in the PNS.

The bar-mode instability can be excited in a hot PNS, a few
milliseconds after core bounce, or, alternatively, it could also be
excited a few tenths of seconds later,
when the PNS cools due to neutrino emission and contracts
further, with $\beta$ becoming larger than the threshold
$\beta_{dyn}$ ( $\beta$ increases roughly as $\sim 1/R$ during
contraction). The amplitude of the emitted
gravitational waves can be estimated as $h\sim M R^2 \Omega^2/d$,
where $M$ is the mass of the body, $R$ its size, $\Omega$ the
rotation rate and $d$ the distance of the source. This leads to an
estimation of the GW amplitude
\begin{equation}
h \approx 9 \times 10^{-23} \left(\frac{\epsilon}{0.2} \right)
\left(\frac{f}{3 {\rm kHz}}\right)^2 \left(\frac{15 {\rm
Mpc}}{d}\right) M_{1.4} R_{10}^2.
\end{equation}
where $\epsilon$ measures the ellipticity of the bar, $M$ is measured
in units of $1.4 M_\odot$ and $R$ is measured in units of 10km.
Notice that, in uniformly rotation Maclaurin spheroids,  the
GW frequency $f$ is twice the rotational frequency $\Omega$. Such a
signal is detectable only from sources in our galaxy or the nearby
ones (our Local Group). If the sensitivity of the detectors is
improved in the kHz region, signals from the Virgo cluster could
be detectable. If the bar persists for many ($\sim$ 10-100) rotation
periods, then even signals from distances considerably larger than
the Virgo cluster will be detectable. Due to the requirement of
rapid rotation, the event rate of the classical dynamical
instability is considerably lower than the SN event rate.

The above estimates rely on Newtonian calculations; GR enhances the
onset of the instability, $\beta_{\rm
dyn}\sim 0.24$ \cite{SBS2000,Saijo01} and somewhat lower than that
for large compactness (large $M/R$). Fully relativistic
dynamical simulations of this instability have been obtained,
including detailed waveforms of the associated gravitational wave
emission. A detailed investigation of the required initial conditions
of the progenitor core, which can lead to the onset of the dynamical
bar-mode instability in the formed PNS, was presented
in\cite{Shibata04}. The amplitude of gravitational waves was due to
the bar-mode
instability was found to be larger by
an order of magnitude, compared to the axisymmetric core
collapse signal.

\vskip 0.4cm

\noindent{\it Low-$T/|W|$ instabilities.}
The {\it bar-mode} instability may be excited for significantly
smaller $\beta$, if centrifugal forces produce a peak in the density
off the source's rotational center\cite{Centrella2001}. Rotating stars
with a high degree of differential rotation are also dynamically
unstable for significantly lower $\beta_{\rm dyn}\gtrsim 0.01$
\cite{Shibata2002,Shibata2003}. According to this scenario the
unstable neutron
star settles down to a non-axisymmetric quasi-stationary state which
is a strong emitter of quasi-periodic gravitational waves
\begin{equation}
h_{\rm eff} \approx 3\times 10^{-22} \left(\frac{R_{\rm eq}}{30 {\rm
km}} \right) \left(\frac{f} {800 {\rm Hz}}\right)^{1/2} \left(\frac{100
{\rm Mpc}} {d}\right) M_{1.4}^{1/2} .
\end{equation}
The bar-mode instability of differentially rotating neutron stars is
an excellent source of gravitational waves, provided the high degree
of differential rotation that is required can be realized. One
should also consider the effects of viscosity and magnetic fields.
If magnetic fields enforce uniform rotation on a short timescale,
this could have strong consequences regarding the appearance and
duration of the dynamical nonaxisymmetric instabilities.

An $m=1$ {\it one-armed spiral} instability has also been shown to
become unstable in proto-neutron stars, provided that the
differential rotation is sufficiently strong
\cite{Centrella2001,SBM2003}. Although it is dominated by a
``dipole" mode, the instability has a spiral character, conserving
the center of mass. The onset of the instability appears to be
linked to the presence of corotation points\cite{Saijo05}
 (a similar link to corotation points has been proposed for
 the low-$T/|W|$ bar mode instability\cite{Watts04}
) and requires a very high degree of differential rotation (with
matter on the axis rotating at least 10 times faster than
matter on the equator). The $m=1$ spiral instability was recently
observed in simulations of rotating core collapse, which started
with the core of an evolved 20$M_\odot$ progenitor star to which
differential rotation was added\cite{Ott05}. Growing from noise level
($\sim 10^{-6}$) on a timescale
of ~5ms, the $m=1$ mode reached its maximum amplitude after
$\sim 100$ms.
Gravitational waves were emitted through the excitation
of an $m=2$ nonlinear harmonic at a frequency of $\sim 800$Hz with
an amplitude comparable to the core-bounce axisymmetric signal.

\subsection{Secular gravitational-wave-driven instabilities}

In a nonrotating star, the forward and backward moving modes
of same $(l,|m|)$ (corresponding to $(l,+m)$ and $(l,-m)$) have
eigenfrequencies $\pm |\sigma|$. Rotation splits this degeneracy
by an amount $\delta \sigma \sim m \Omega$ and both the
prograde and retrograde modes are dragged forward by the stellar
rotation. If the star spins sufficiently rapidly, a
mode which is retrograde (in the frame rotating with the star)
will appear as prograde in the inertial frame (a nonrotating observer
at infinity). Thus, an inertial observer sees GWs with positive
angular momentum emitted by the retrograde mode, but since the
perturbed fluid rotates slower than it would in the absence of the
perturbation, the angular momentum of the mode in the rotating
frame is negative. The emission of
GWs consequently makes the angular momentum of the mode increasingly
negative, leading to the instability. A mode is unstable when
$\sigma(\sigma-m\Omega) < 0$.
This class of {\em frame-dragging instabilities} is usually referred
to as Chandrasekhar-Friedman-Schutz\cite{Chandra70,Friedman78} (CFS)
instabilities.

\vskip 0.4cm

\noindent{\it $f$-mode instability.}
In the Newtonian limit, the $l=m=2$ $f$-mode (which has the shortest
growth time of all polar fluid modes) becomes unstable when
$T/|W|>0.14$, which is near or even above the mass-shedding
limit for typical polytropic EOSs used to model uniformly rotating
neutron stars. Dissipative effects (e.g. shear and bulk viscosity or
mutual friction in
superfluids)\cite{Cutler87,Lindblom79,Ipser91,Lindblom95}
leave only a small instability window near mass-shedding, at
temperatures of $\sim 10^9$K. However,
relativistic effects strengthen the instability considerably, lowering
the required $\beta$ to  $\approx 0.06-0.08$ \cite{SF1998,MSB99}
for most realistic EOSs and masses of $\sim 1.4M_\odot$ (for higher
masses, such as hypermassive stars created in a binary NS merger,
the required rotation rates are even lower).

Since PNSs rotate differentially, the above limits derived under
the assumption of uniform rotation are too strict. Unless uniform
rotation is enforced on a short timescale, due to e.g. magnetic
braking\cite{Liu04},
the $f$-mode instability will develop in a differentially rotating
background, in which the required $T/|W|$ is only somewhat larger
than the corresponding value for uniform rotation\cite{Yoshida}, but
the mass-shedding limit is dramatically relaxed. Thus, in a
differentially
rotating PNS, the $f$-mode instability window is huge, compared to
the case of uniform rotation and the instability can develop provided
there is sufficient $T/|W|$ to begin with.

The $f$-mode instability is an excellent source of GWs.
Simulations of its nonlinear development in the ellipsoidal
approximation\cite{Lai95}
have shown that the mode can grow to a large nonlinear amplitude,
modifying the background star from an axisymmetric shape to a
differentially rotating ellipsoid. In this modified background the
$f$-mode amplitude saturates and the ellipsoid becomes a strong
emitter of gravitational waves, radiating away angular momentum
until the star is slowed-down towards a stationary state. In the
case of uniform density ellipsoids, this stationary state is the
Dedekind ellipsoid, i.e. a nonaxisymmetric ellipsoid with internal
flows but with a stationary (nonradiating) shape in the inertial
frame. In the ellipsoidal approximation, the nonaxisymmetric
pattern radiates gravitational waves sweeping through the LIGO II
sensitivity window (from 1kHz down to about 100Hz) which could
become detectable out to a distance of more than 100Mpc.

Two recent hydrodynamical simulations\cite{ShibataKarino04,Ou04}
 (in the Newtonian limit
and using a post-Newtonian radiation-reaction potential)
essentially confirm this picture. In {}\cite{ShibataKarino04}
  a differentially rotating, $N=1$ polytropic model with a
 large $T/|W| \sim 0.2-0.26$ is chosen as the initial equilibrium
 state. The main difference
of this simulation compared to the ellipsoidal approximation
comes from the choice of EOS. For $N=1$ Newtonian polytropes
it is argued that the secular evolution cannot lead to a
a stationary Dedekind-like state does not exist. Instead,
the $f$-mode instability will continue to be
active until all nonaxisymmetries are radiated away and an
axisymmetric shape is reached. This conclusion should be
checked when relativistic effects are taken into account,
since, contrary to the Newtonian case, relativistic $N=1$
uniformly rotating polytropes {\it are} unstable to the $l=m=2$
$f$-mode\cite{SF1998}
 -- however it has not become possible, to date, to
construct relativistic analogs of Dedekind ellipsoids.

In the other recent simulation {}\cite{Ou04}
, the initial state was chosen to be a uniformly rotating, $N=0.5$
polytropic model with $T/|W|\sim 0.18$. Again, the main conclusions
reached in {}\cite{Lai95}
 are confirmed, however, the assumption of uniform
initial rotation limits the available angular momentum that can be
radiated away, leading to a detectable signal only out to about
$\sim 40$Mpc. The star appears to be driven towards a Dedekind-like
state, but after about 10 dynamical periods, the shape is disrupted
by growing short-wavelength motions, which are suggested to arise
because of a shearing type instability, such as the elliptic flow
instability\cite{Lifschitz93}.

\vskip 0.4cm

\noindent{\it $r$-mode instability.} Rotation does not only shift
the spectra of polar modes; it also lifts the degeneracy of axial
modes, give rise to a new family of {\em inertial} modes, of which
the $l=m=2$ $r$-mode is a special member. The restoring force, for
these oscillations is the Coriolis force. Inertial modes are
primarily velocity perturbations. The frequency of the $r$-mode in
the rotating frame of reference is $\sigma = 2 \Omega /3$. According
to the criterion for the onset of the CFS instability, the $r$-mode
is unstable for any rotation rate of the
star\cite{Andersson98,FriedmanMorsink}. For temperatures between
$10^{7}-10^{9}$K and rotation rates larger than 5-10\% of the Kepler
limit, the growth time of the unstable mode is smaller than the
damping times of the bulk and shear viscosity\cite{LOM98,AKS99}. The
existence of a solid crust  or of hyperons in the core \cite{LO2002}
 and magnetic fields \cite{Rezzolla1,Rezzolla2}, can also significantly affect the
onset of he instability (for extended reviews see
\cite{AK2001,Nils2003}). The suppression of the $r$-mode instability
by the presence of hyperons in the core is not expected to operate
efficiently in rapidly rotating stars, since the central density is
probably too low to allow for hyperon formation. Moreover, a recent
calculation\cite{vanDalen03} finds the contribution of hyperons to
the bulk viscosity to be two orders of magnitude smaller than
previously estimated. If accreting neutron stars in Low Mass X-Ray
Binaries (LMXB, considered to be the progenitors of millisecond
pulsars) are shown to reach high masses of $\sim 1.8M_\odot$, then
the EOS could be too stiff to allow for hyperons in the core (for
recent observations that support a high mass for some millisecond
pulsars see {}\cite{Nice03}).

The unstable $r$-mode grows exponentially until it saturates due to
nonlinear effects at some maximum amplitude $\alpha_{max}$. The
first computation of nonlinear mode couplings using second-order
perturbation theory suggested that the $r$-mode is limited to very
small amplitudes (of order $10^{-3}-10^{-4}$) due to transfer of
energy to a large number of other inertial modes, in the form of a
cascade, leading to an equilibrium distribution of mode
amplitudes\cite{Arras2002}. The small saturation values for the
amplitude are supported by recent nonlinear estimations
\cite{Sa2004,SaTome2005} based on the drift, induced by the r-modes,
causing differential rotation.  On the other hand, hydrodynamical
simulations of limited resolution showed that an initially
large-amplitude $r$-mode does not decay appreciably over several
dynamical timescales\cite{StergioulasFont} , but on a somewhat
longer timescale a catastrophic decay was observed\cite{Gressman02}
 indicating a transfer of energy to other modes, due to nonlinear
 mode couplings and suggesting that a hydrodynamical instability may
 be operating. A specific resonant 3-mode coupling was
identified\cite{Lin04} as
the cause of the instability and a perturbative analysis of the decay
rate suggests a maximum saturation amplitude $\alpha_{max} < 10^{-2}$.
A new computation using second-order perturbation
 theory finds that the catastrophic decay seen in the hydrodynamical
 simulations\cite{Gressman02,Lin04}
 can indeed be explained by a parametric instability operating
 in 3-mode couplings between the $r$-mode and two other inertial
 modes\cite{Brink04a,Brink04b,Brink05}. Whether the maximum saturation
amplitude is set by a network
 of 3-mode couplings or a cascade is reached, is, however, still
 unclear.

A neutron star spinning down due to the $r$-mode instability
will emit gravitational waves of amplitude
\begin{equation}
h(t)\approx 10^{-21} \alpha \left( \frac{\Omega}{1 {\rm
kHz}}\right)\left(\frac{100 {\rm kpc}}{d}\right)
\end{equation}
Since $\alpha$ is small, even with LIGO II the signal is
undetectable at large distances (VIRGO cluster) where the SN
event rate is appreciable, but could be detectable after
long-time integration from a galactic event. However, if
the compact object is a strange star, then the instability may
not reach high amplitudes ($\alpha \sim 10^{-3}-10^{-4}$) but it
will persist for a few hundred years (due to the different
temperature dependence of viscosity in strange quark matter)
and in this case there might be
up to ten unstable stars in our galaxy at any time \cite{AJK2002}.
Integrating data for a few weeks could lead to an effective amplitude
$h_{\rm eff}\sim 10^{-21}$ for galactic signals at frequencies $\sim
700-1000$Hz. The frequency of the signal changes only slightly on a
timescale of a few months, so that the radiation is practically
monochromatic.

\vskip 0.4cm

\noindent{\it Other unstable modes.}  The CFS instability can also
operate for core g-mode oscillations\cite{Lai99}
 but also for {\em w}-mode oscillations, which are basically spacetime
 modes\cite{KRA2004}. In addition, the CFS instability can operate
 through other dissipative effects. Instead of the gravitational
radiation, any radiative mechanism (such as electromagnetic
radiation) can in principle lead to an instability.

\subsection{Secular viscosity-driven instability}

A different type of nonaxisymmetric instability in rotating stars is
the instability driven by viscosity, which breaks the circulation of
the fluid \cite{RS63,Ja64}. The instability is suppressed by
gravitational radiation, so it cannot act in the temperature window in
which the CFS-instability is active. The instability
sets in when the frequency of a {\it prograde} $l=-m$ mode goes
through zero in the
rotating frame. In contrast to the CFS-instability, the
viscosity-driven instability is not generic in rotating stars. The
$m=2$ mode becomes unstable at a high rotation rate for very stiff
stars and higher $m$-modes become unstable at {\it larger} rotation
rates.

In Newtonian polytropes, the instability occurs only for stiff
polytropes of index $N<0.808$ \cite{Ja64,SL96}. For relativistic
models, the situation for the instability becomes worse, since
relativistic effects tend to suppress the viscosity-driven instability
(while the CFS-instability becomes stronger). For the most
relativistic stars, the viscosity-driven bar mode can become unstable
only if $N<0.55$\cite{BFG97}.  For $1.4 M_{\odot}$ stars, the
instability is present for $N<0.67$.

An investigation of the viscosity-driven bar mode instability, using
incompressible, uniformly rotating triaxial ellipsoids in the
post-Newtonian approximation\cite{SZ97} finds
that the relativistic effects increase the critical $T/|W|$ ratio for
the onset of the instability significantly. More recently, new
post-Newtonian \cite{DiGirolamo02} and fully relativistic calculations
for uniform-density stars \cite{Gondek02} show that the
viscosity-driven instability is not as strongly suppressed by
relativistic effects as suggested in \cite{SZ97}. The most promising
case for the onset of the viscosity-driven instability (in terms of
the critical rotation rate) would be rapidly rotating strange stars
\cite{Gondek03}, but the instability can only appear if its growth
rate is larger than the damping rate due to the emission of
gravitational radiation - a corresponding detailed comparison is still
missing.

\section{Accreting neutron stars in LMXBs}

Spinning neutron stars with even tiny deformations are interesting
sources of gravitational waves. The deformations might results from
various factors but it seems that the most interesting cases are the
ones in which the deformations are caused by accreting material. A
class of objects called Low-Mass X-Ray Binaries (LMXB) consist of a
fast rotating neutron star (spin $\approx 270-650$Hz) torqued by
accreting material from a companion star which has filled up its
Roche lobe. The material adds both mass and angular momentum to the
star, which, on timescales of the order of tenths of Megayears
could, in principle, spin up the neutron star to its break up limit.
One viable scenario \cite{Bildsten1998} suggests that the accreted
material (mainly hydrogen and helium) after an initial phase of
thermonuclear burning undergoes a non-uniform crystallization,
forming a crust at densities $\sim 10^8-10^9$g/cm$^3$. The
quadrupole moment of the deformed crust is the source of the emitted
gravitational radiation which slows-down the star, or halts the
spin-up by accretion.

An alternative scenario has been proposed by
Wagoner\cite{Wagoner1984} as a follow up of an earlier idea by
Papaloizou-Pringle\cite{PP1978}. The suggestion was that the spin-up
due to accretion might excite the $f$-mode instability, before the
rotation reaches the breakup spin. The emission of gravitational
waves will torque down the star's spin at the same rate as the
accretion will torque it up, however, it is questionable whether the
$f$-mode instability will ever be excited for old, accreting neutron
stars. Following the discovery that the $r$-modes are unstable at
any rotation rate, this scenario has been revived independently by
Bildsten \cite{Bildsten1998} and Andersson, Kokkotas and Stergioulas
\cite{AKS1999}. The amplitude of the emitted gravitational waves
from such a process is quite small, even for high accretion rates,
but the sources are persistent and in our galactic neighborhood the
expected amplitude is
\begin{equation}
h\approx 10^{-27}\left(\frac{1.6\mbox{ms}}{P}
\right)^5\frac{1.5\mbox{kpc}}{D} \, .
\end{equation}
This signal is within reach of advanced LIGO with signal recycling
tuned at the appropriate frequency and integrating for a few
months\cite{ThorneCutler}. This picture is in practice more
complicated, since the growth rate of the $r$-modes (and
consequently the rate of gravitational wave emission) is a function
of the core temperature of the star. This leads to a thermal runaway
due to the heat released as viscous damping mechanisms counteract
the r-mode growth \cite{Levin99}. Thus, the system executes a limit
cycle, spinning up for several million years and spinning down in a
much shorter period. The duration of the unstable part of the cycle
depends critically on the saturation amplitude $\alpha_{max}$ of the
$r$-modes \cite{AJKS2000,Heyl}. Since current computations
\cite{Arras2002,SaTome2005} suggest an $\alpha_{max}\sim
10^{-3}-10^{-4}$, this leads to a quite long duration for the
unstable part of the cycle of the order of $\sim 1Myear$.

The instability window depends critically on the effect of the shear
and bulk viscosity and various alternative scenarios might be
considered. The existence of hyperons in the core of neutron stars
induces much stronger bulk viscosity which suggests a much narrower
instability window for the $r$-modes and the bulk viscosity prevails
over the instability even in temperatures as low as $10^8$K
\cite{LO2002}. A similar picture can be drawn if the star is
composed of ``deconfined" $u$, $d$ and $s$ quarks - a strange star
\cite{Madsen98}. In this case, there is a possibility that the
strange stars in LMXBs evolve into a quasi-steady state with nearly
constant rotation rate, temperature and mode amplitude
\cite{AJK2002} emitting gravitational waves for as long as the
accretion lasts. This result has also been found later for stars
with hyperon cores \cite{Wagoner,Reisenegger03}. It is interesting
that the stalling of the spin up in millisecond pulsars (MSPs) due
to $r$-modes is in good agreement with the minimum observed period
and the clustering of the frequencies of MSPs \cite{AJKS2000}.

\section{Binary mergers}

Depending on the high-density EOS and their initial masses, the
outcome of the merger of two neutron stars may not always be a black
hole, but a hypermassive, differentially rotating compact star (even
if it is only temporarily supported against collapse by differential
rotation). A recent detailed simulation\cite{Shibata05b} in full GR
has shown that the hypermassive object created in a binary NS merger
is nonaxisymmetric. The nonaxisymmetry lasts for a large number of
rotational periods, leading to the emission of gravitational waves
with a frequency of $~3$kHz and an effective amplitude of $\sim
6-7\times 10^{-21}$ at a large distance of 50Mpc. Such large
effective amplitude may be detectable even by LIGO II at this high
frequency.

The tidal disruption of a NS by a BH \cite{Vallisneri} or the merging
of two NSs \cite{Rasio} may give valuable information for the radius
and the EoS if we can recover the signal at frequencies higher than 1
kHz.

\section{Gravitational-wave asteroseismology}

If various types of oscillation modes are excited during the
formation of a compact star and become detectable by gravitational
wave emission, one could try to identify observed frequencies with
frequencies obtained by mode-calculations for a wide parameter range
of masses, angular momenta and EOSs.
\cite{AK1998,Benhar1999,KAA2001,Sotani2004a,Sotani2004b,Sotani2005,Benhar2004}.
Thus, {\em gravitational wave asteroseismology} could enable us to
estimate the mass, radius and rotation rate of compact stars,
leading to the determination of the "best-candidate" high-density
EoS, which is still very uncertain. For this to happen, accurate
frequencies for different mode-sequences of rapidly rotating compact
objects have to be computed.

For slowly rotating stars, the frequencies of $f-$, $p-$ and $w-$
modes are still unaffected by rotation, and one can construct
approximate formulae in order to relate observed frequencies and
damping times of the various stellar modes to stellar parameters. For
example, for the fundamental oscillation ($l=2$) mode ($f$-mode) of
non-rotating stars one obtains \cite{AK1998}
\begin{eqnarray}
\sigma({\rm kHz})&\approx& 0.8+1.6 M_{1.4}^{1/2}R_{10}^{-3/2}
+ \delta_1 m{\bar \Omega} \\
\tau^{-1}({\rm secs}^{-1})&\approx&
M_{1.4}^3R_{10}^{-4}\left(22.9-14.7M_{1.4}R_{10}^{-1}\right)+
\delta_2 m {\bar \Omega}
\end{eqnarray}
where ${\bar \Omega}$ is the normalized rotation frequency of the
star, and $\delta_1$ and $\delta_2$ are constants estimated by
sampling data from various EOSs. The typical frequencies of NS
oscillation modes are larger than 1kHz. Since each type of mode is
sensitive to the physical conditions where the amplitude of the mode
is largest, the more oscillations modes can be identified through
gravitation waves, the better we will understand the detailed
internal structure of compact objects, such as the existence
of a possible superfluid state of matter\cite{Andersson01}.

If, on the other hand, some compact stars are born rapidly rotating
with moderate differential rotation, then their central densities
will be much smaller than the central density of a nonrotating star
or same baryonic mass. Correspondingly, the typical axisymmetric
oscillation frequencies will be smaller than 1kHz, which is more
favorable for the sensitivity window of current interferometric
detectors\cite{SAF}. Indeed, axisymmetric simulations of rotating
core-collapse have shown that if a rapidly rotating NS is created,
then the dominant frequency of the core-bounce signal (originating
from the fundamental $l=2$ mode or the $l=2$ piece of the
fundamental quasi-radial mode) is in the range
600Hz-1kHz\cite{DFM2002b}.

If different type of signals are observed after core collapse,
such as both an axisymmetric core-bounce signal and a
nonaxisymmetric one-armed instability signal, with a time
separation of the order of 100ms, this would yield
invaluable information about the angular momentum distribution
in the proto-neutron stars.


\end{document}